\documentstyle[psfig]{l-aa}

\begin{document}

   \thesaurus{06         % A&A Section 6: Form. struct. and evolut. of stars
              (03.11.1;  % Cosmogony,
               16.06.1;  % Planets and satellites: general,
               19.06.1;  % Solar system: general,
               19.37.1;  % Stars: formation of,
               19.53.1;  % Stars: oscillations of,
               19.63.1)} % Stars: structure of.
   \title{The Dark and Luminous Matter coupling in the 
formation of spheroids:  a SPH investigation}
   \author{Cesario Lia
          \inst{1}, Giovanni Carraro\inst{2} and Paolo Salucci
          \inst{1}
          }

   \offprints{G. Carraro ({\tt carraro@pd.astro.it})}

   \institute{SISSA/ISAS, via Beirut 2, I-34013, Trieste,
	Italy
        \and
         Department of Astronomy, Padova University,
	vicolo dell'Osservatorio 5, I-35122, Padova, Italy\\
	    e-mail: {\tt
liac,salucci\char64sissa.it,carraro\char64pd.astro.it}
             }

   \date{Received February 16; accepted ......}

   \maketitle

   \markboth{Lia, Carraro \& Salucci}{Dark and Luminous Matter Coupling}

   \begin{abstract}

Using N--body/hydrodynamical simulations
which include prescriptions for Star Formation, Feed-Back and
Chemical Evolution, we explore the interaction
between baryons and Dark Matter (DM)
at galactic scale.\\
The N--body simulations are performed using a Tree--SPH code that
follows
the evolution of individual DM halos inside which stars form from cooling
gas, and evolve delivering in the interstellar medium (ISM) mass, metals,
 and energy.\\
We examine the formation and evolution of a giant and
a dwarf elliptical galaxy of total mass $10^{12} M_{\odot}$ and $10^{9}
M_{\odot}$, respectively.\\
Starting from an initial density profile like  the   
universal  
Navarro et al (1996) profile in the inner region, baryons sink towards
the  center
due to  cooling energy losses.
At the end of the collapse, the innermost part ($\simeq 1/20$ of the halo size) of the galaxy is
baryon-dominated, whereas the outer regions are DM dominated.\\ 
The star  formation proceeds at a much faster speed in the giant galaxy where
a spheroid of  $8\times 10^{10}M_\odot$
is formed in $2~ Gyr$,  with respect to the dwarf galaxy where
the spheroid  of  $2\times  10^7M_\odot $ is formed in $4~Gyr$.
 For the two objects  the final
distributions of stars  are well fitted by a Hernquist profile with effective radii of $r_e=30$
kpc and $2.8$ kpc, respectively.
The dark-to-luminous transition radius $r_{IBD}$ occurs roughly at $1~r_e$, as in 
real ellipticals.
The  DM halo density  evolution is non-adiabatic and does not  lead to   a core
radius.

   \end{abstract}

%
%  14.Sep.'90: Demo-Vs.
%________________________________________________________________

\section{Introduction}

Galaxies of any morphological type and luminosity are known to be
surrounded by DM halos, whose properties are remarkably universal (Salucci
\& Persic 1997 and references therein).
The presence of DM halos has been detected through a variety of
observational methods (Danziger 1997), going from rotation curves in
spirals (Giraud 2000, Swaters 1999, Persic et al 1996)  
to $M/L$ ratios in ellipticals (Bertola et al
1993, Loewenstein \& White III 1999 and references therein).
To summarize, the properties of DM halos can be  described as follows
(Salucci \& Persic 1997):

\begin{description}
\item $\bullet$ dark
and visible matter are well mixed already inside the
luminous region of the galaxy;
\item $\bullet$ the transition radius $R_{IBD}$ between the inner, baryon dominated region,
and the outer, DM dominated region, moves inward progressively with
decreasing luminosity;
\item $\bullet$ a halo core radius, comparable with the optical radius, is
detected
at all luminosities and for all morphologies;
\item $\bullet$ the luminous mass fraction varies with luminosity in a
fashion common to all galaxy types: it is comparable with the cosmological
baryon fraction at $L \simeq  L_{\star}$, but it decreases by about  a factor
100  at $L << L_{\star}$;
\item $\bullet$ finally, for any Hubble type, the central halo density
increases with decreasing luminosity. 
\end{description}

%%%%%%%%%%%%%%%%%%%Figure 1
\begin{figure*}
\centerline{\psfig{file=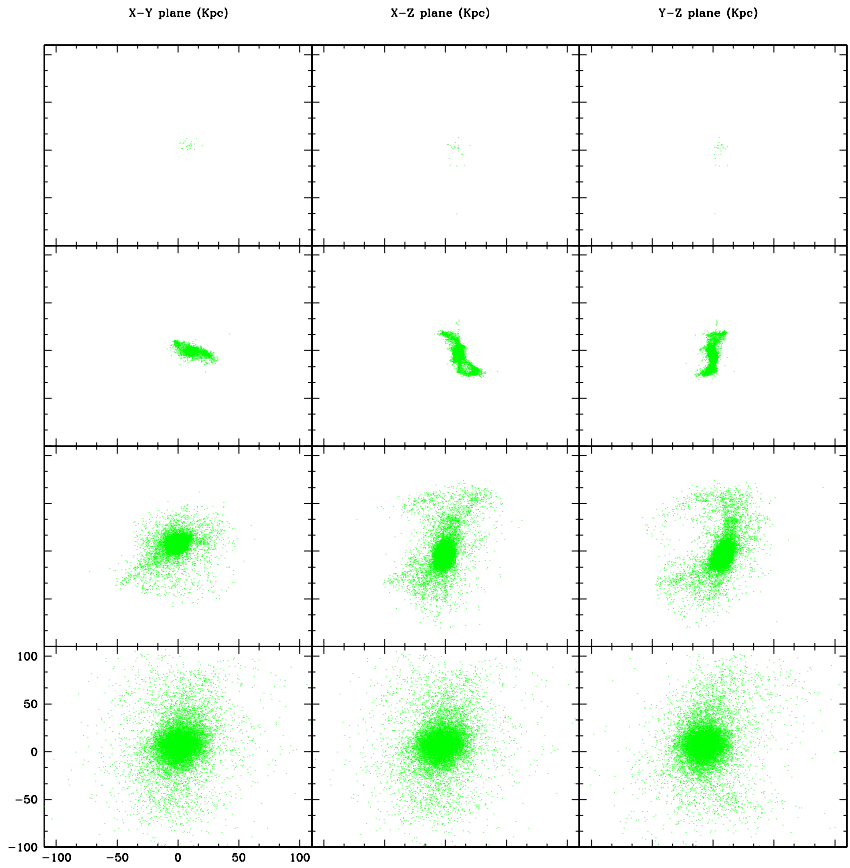,height=14cm,width=14cm}}
\caption{The formation of a giant elliptical. 
From the top to the bottom,
snapshots refer to 1, 2, 3 and 9 Gyrs. }
\end{figure*}

 %%%%%%%%%%%%%%%%%%%Figure 2
\begin{figure}
\centerline{\psfig{file=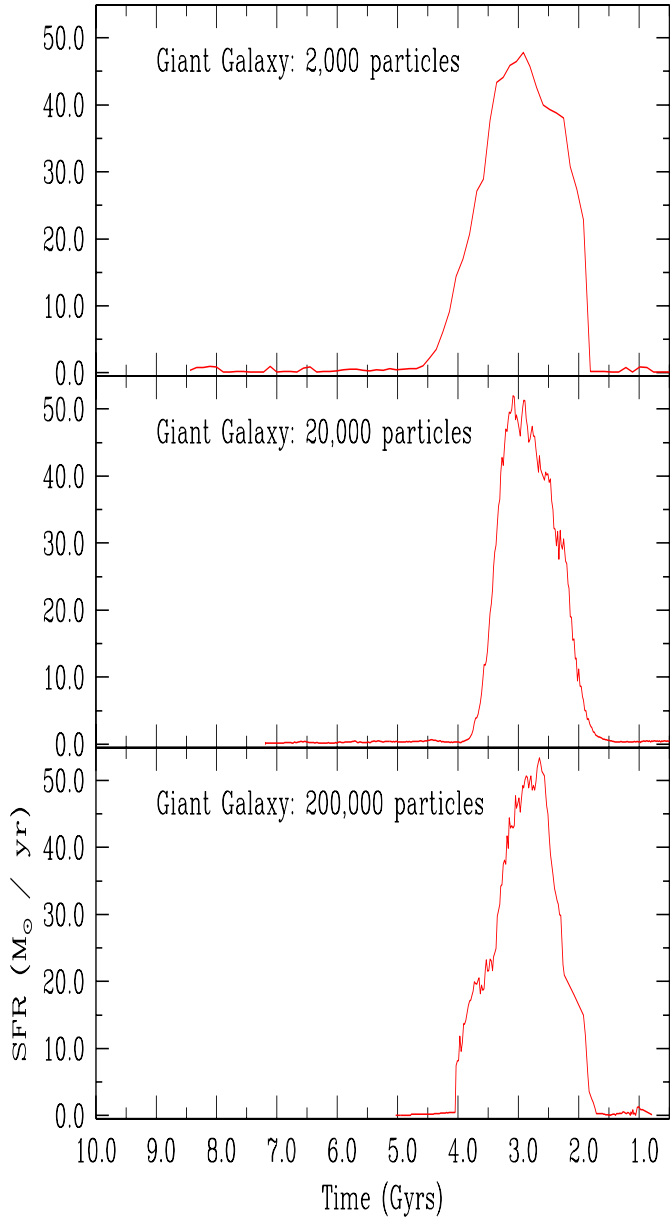,height=18cm,width=10cm}}
\caption{SF for the giant elliptical as a function of time. From the top to the
bottom the same model is shown at increasing number of particles, i.e at
increasing resolution.}
\end{figure}

%%%%%%%%%%%%%%%%%%%%Figure 3
\begin{figure}
\centerline{\psfig{file=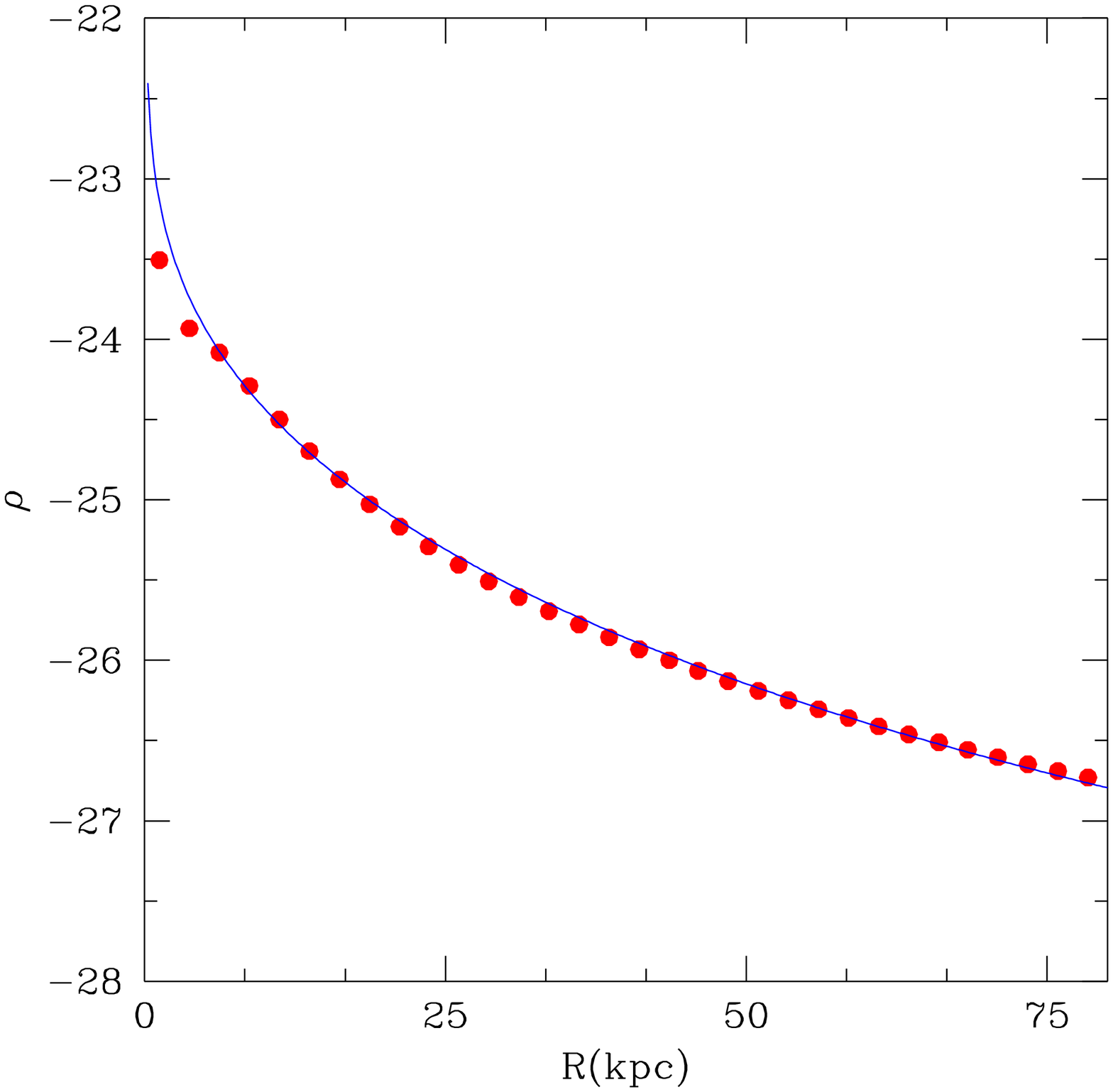,height=9cm,width=9cm}}
\caption{Final stellar density profile of the giant galaxy. Overimposed
is a Hernquist profile for $r_e~=~30$ kpc.}
\end{figure}

Attempts to model the  properties of DM halos in a cosmological context with N--body
simulations trace back to Dubinski \& Carlberg (1991) in the frame  of  
Cold Dark Matter (CDM) theory.  The halos
were found to be strongly triaxial and to exhibit a power law density
profile varying from $-1$ in the center to $-4$ in the outskirts
(Hernquist 1990 profile).
Then,   Navarro et al (1996) 
(hereafter NFW), found that, independently from the adopted initial
perturbation spectrum, the  cosmological model and the halo mass,
all DM halos possess the same {\it universal} density profile, fitted by
the formula

\begin{equation}
\frac{\rho_{r}}{\rho_{crit}}  = \frac{\delta_{c}}{(r/r_{s})^a(1+ r/r_{s})^{3-a}}  ,
\end{equation}

\noindent
where $a=1$, $\rho_{crit} = 3 H_{o}^{2}/8 \pi G$ is the critical density for closure,
$\delta_{c}$ is a dimensionless characteristic density, and $r_{s}$ is a scale
radius, which defines where the profile shape has slope
$-2$. \footnote{ $^*$ We take as 
Hubble's constant $H_{o} = 75~Km/sec/Mpc$.}\\

This profile has a spike in the center of the halo, and 
differs  in its asymptotic behavior from the Hernquist
profile, decreasing as $r^{-3}$ far from the halo center.
Other simulations, with higher resolution and/or different initial
conditions, confirmed the basic features of 
these findings, but disagreed in some important  aspects  (Cole \& Lacey 1996, Moore et al
1997, van der Bosh, 1999). In fact,  there are claims  that the  {\it universality} 
of the functional form (1)  arises as a direct consequence of  the
hierarchical merging history of CDM halos (Syer \& White (1997),  or as a  more
generic feature of gravitational collapse (Huss et al.  1998). However,  recently,   the   
steepness of the central  cusp  has been found to significantly vary
among different realizations, i.e.  among halos (Jing \& Suto 2000).  It seems now likely that,  CDM halos 
follow   eq (1)  with   $a \simeq 1.5$ but with  large variations of $r_s$ with mass and also at
a given  halo  mass.\\

On the other hand, there exists  a large discrepancy between  CDM halo 
predictions  and DM 
observations (Salucci \& Persic 1997). Halos around galaxies show a density distribution inconsistent
with eq (1). In particular,  they have a  density central  core    larger than  the
stellar scale-length and their density is:
$$
\rho_h(r)\propto {1\over{(r+r_0)(r_0^2+r^2)}}
$$
with $r_0>>r_e$, $r_e$ being the effective radius.

The   disagreement between theory 
and observations on the mass distribution,  
 and  the existence of global scaling laws
that couple the dark and the luminous matter (Persic et al 1996)
 prompt  the   investigation 
of the past dynamical history of galaxies.  

N-body/hydrodynamical simulations  are an effective tool
to  obtain  crucial  information on the late stages of galaxy formation in some
sense orthogonal 
  to that  we obtain  with
semi-analitycal methods or we infer   from observations.
In fact,  such simulations can  account for  
the "physical"  interaction between gas and dark matter .
Moreover,  many relevant  physical processes occurring in
the baryonic components, like thermal shocks, pressure forces
and  dissipation are explicitly taken into account.

The layout of the paper is as follow. In Section~2 we briefly
describe the  numerical tool, in Section~3 we discuss 
the initial conditions.
In Section~4 and 5 we show the evolution of a giant and a dwarf elliptical,
respectively. Finally, Section~6 summarizes the results.

\section{The CODE}
The simulations  we present  here  have been performed
by means of  the Tree--SPH
code developed by Carraro (et al 1998),  Buonomo et al (2000)
 and Lia \& Carraro (2000).
The code, which is able to  follow the evolution of a mix of  CDM 
 and Baryons (gas and stars),
 has been successfully  checked against  standard tests in Carraro et al. (1998),
 while a fine 
 exploration of the parameters space is presented in Buonomo et al (2000).

In detail, the gas component is 
"followed"  by means of the Smoothed Particle
Hydrodynamics (SPH) technique (Lucy 1977; 
Gingold \& Monaghan 1997; Hernquist \& Katz 1989; Steinmetz \& M\"uller 1993), while  the
gravitational forces are taken into account 
by means of the hierarchical tree algorithm of Barnes \& Hut (1986). In detail,  we adopt 
a tolerance parameter $\theta=0.8$, a Plummer softening parameter and we expand the  tree nodes to
quadrupole order.

In SPH each  particle represents a fluid element whose
position, velocity, energy, density etc. are followed in  time and space.
The properties of the fluid are locally estimated by an interpolation
which
involves the smoothing length $h_{i}$.
 Each particle possesses its own time and space
variable smoothing
length $h_{i}$, and evolves with its own time-step.
This renders the code highly adaptive and flexible, and
suited for  numerical "experiments".

%%%%%%%%%%%%%%%%%%%%Figure 4
 \begin{figure*}
 \centerline{\psfig{file=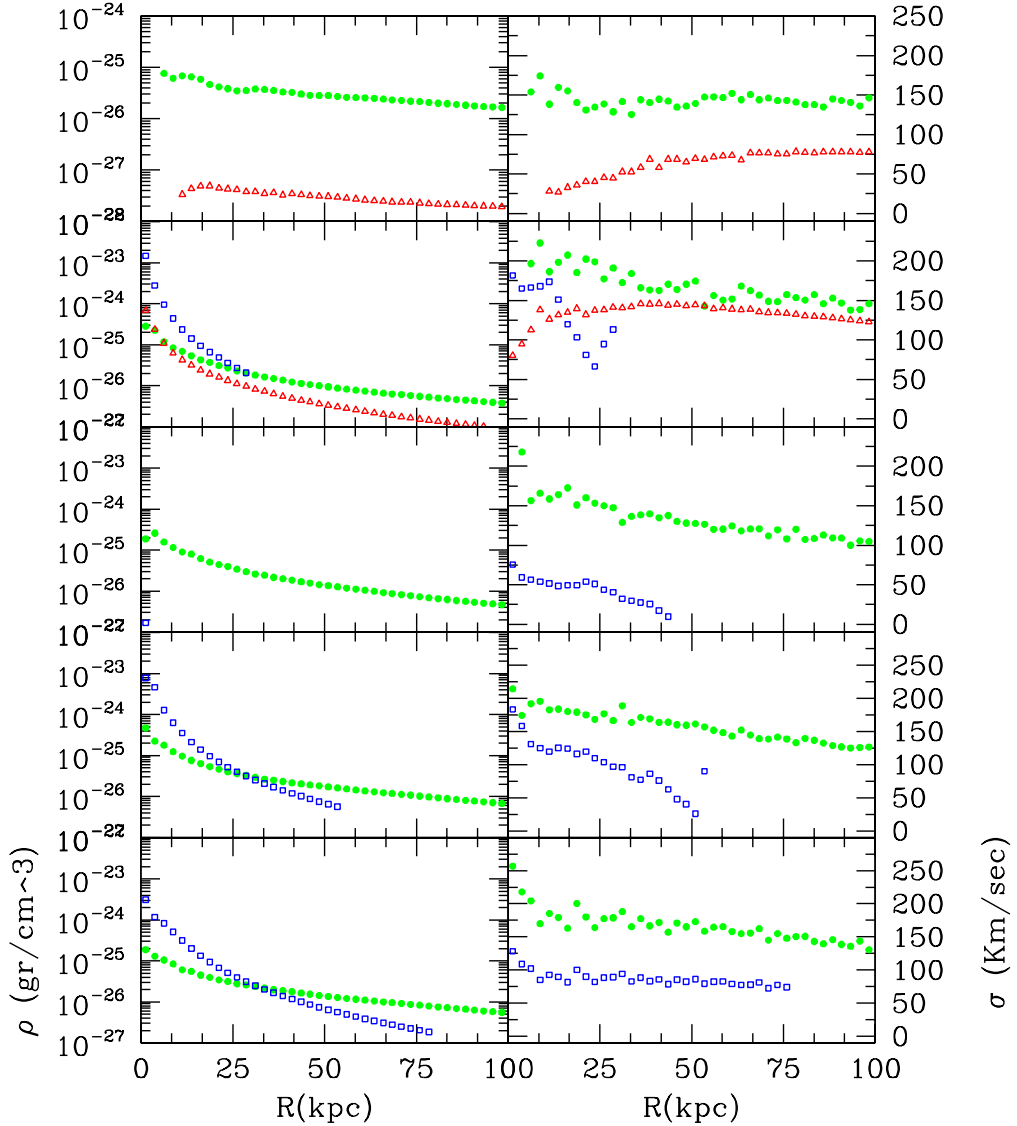,height=16cm,width=16cm}}
 \caption{Time evolution of the  density and dispersion profiles. Solid circles stand for DM, open triangles for
 gas and open squares for stars. From the top to
 the bottom profiles refer to 1, 3, 3.5, 4.5 and 9 Gyr.}
 \end{figure*}

Radiative cooling is  considered  as a
function of temperature and metallicity  following 
Sutherland \& Dopita (1994) and Hollenbach \& McKee (1979) : 
the code takes  into account  the variations
in metallicity among the  fluid elements 
as function of time and position.\\
Star Formation (SF) and Feed-back algorithms are described in
Buonomo et al (2000). Specifically  SF, following partly Katz (1992),
is set to occur when 

$$
t_{sound }> t_{ff}
$$
 and
$$
 t_{cooling} << t_{ff}
$$

with the star formation rate of
$$
SFR = \frac{d\rho_{*}}{dt} = -\frac{d\rho_g}{dt} = \frac{c_{*} \rho_g}{t_g}
$$

\noindent
where $c_*$ is the dimensionless efficiency of star formation,
and $t_g$ is the characteristic time for the gas to 
flow, usually set to the maximum between the cooling time and
the free-fall time. For the simulations here discussed we keep $c_* = 1.0$. 
When formed, stars are distributed in mass according to the Miller \& Scalo (1979) 
Initial Mass Function (IMF).\\
The effects of energy
(and mass) feed-back from 
supernovae and stellar winds
are also taken into account (Chiosi \& Maeder 1986, Thornton et al 1998). 
In this experiments we deposit all the energy released by SN\ae~ ($10^{49} erg$)
and stellar wind in the thermal budget of the bubble, following the kind
of arguments discussed in Buonomo et al (2000).

Finally,
the chemical enrichment  of the interstellar  gas  caused by the
the stellar  wind and   ejecta is followed   by means of the  closed-box model applied to
each gas-particle as in  Carraro et al. (1998). Metals are then SPH-diluited over
the surrounding gas particles.

\section{Initial conditions}
 The  code  follows the dynamical  formation of galaxies  since the moment in which 
the protohalo detaches itself from 
the Hubble flow  with its  cosmological share of
baryonic material.   Then, we set {\it ad hoc} initial configuration for a
protogalaxy   and let the system evolve within  the  SPH scheme above described.
Moreover,  we  assume that the star formation starts  after  the dark halo is virialized.
 
This approach  is justified on the grounds that 
in this paper,  we are aimed  at 1)  testing a simple  but reasonable initial conditions set-up 
and 2) 
focusing on processes which occur mainly at scales much smaller  than 
the halo virial radius.  

The  working scenario is that,  after a violent relaxation process, which follows 
the separation from
the Hubble expansion, the  DM halo acquires a  particular density 
distribution, and then  it accretes baryonic
material which heats up to the halo virial temperature.
Gas then radiatively cools and collapses, and through fragmentation turns into the
stars we see today in disks and spheroidal systems.

This scenario is not incompatible with the merging one, withstanding that the baryonic
collapse takes place after the last merging episode.

\section{Initial configuration}
We set up a protogalaxy as an 
isolated virialized DM halo with baryonic material inside it.  
We assume spherical,  isotropic  and  non-rotating  halos of  density
profiles:

\begin{equation}
\rho(r) \propto \frac{1}{r} .
\end{equation}

\noindent

 Let us notice that eq.~(2)    matches   the profile of CDM halos in the
innermost regions, especially in   low $\Omega$ Universe for which 
$r_s \sim 10-30$ kpc. Different profiles, more in line with observations or theoretical 
claims, will be  considered in forthcoming papers.

%%%%%%%%%%%%%%%%%%%%Figure 5
%\begin{figure}
%\centerline{\psfig{file=fig5.ps,height=9cm,width=9cm}}
%\caption{Adiabatic invariant evolution for a giant elliptical in three
%different galactic region. $r \times M(r)$ is in code units.}
%\end{figure}

 Gravitational interaction is modeled by   
adopting  a Plummer softening constant over   the simulation and
 equal for both DM and gas.
By plotting  the inter-particles separation as a function of the
galactocentric distance, we derive the  softening parameter $\epsilon$ as the mean inter-particles
separation at the
center of the sphere, taking  at least one hundred particles
inside the softening radius.   

DM particles (10,000 in number) are distributed inside a  sphere according to an
{\it acceptance--rejection} procedure 
for generating random deviations with a known distribution
function (Press et al 1989).

Along with positions, we assign also the isotropic dispersion
velocity $\sigma(r)$ according to:

\begin{equation}
\sigma^2(r) \propto   (r ln (\frac{1}{r})) .
\end{equation}
\noindent
which is the solution of the Jeans equation for spherical isotropic
collisionless system  with the density profile of eq (2).

Fixing the total mass of the system $M_{200}$, we assume that 
a  radius 

\[
R_{200} \equiv  (\frac{3}{4 \pi})^{1/3} (\frac{M_{200}}{200
\rho_{c}})^{1/3} >> r_e
\]

\noindent
truncates the dark halo, 
where $\rho_{c}$ is the critical density to close the Universe.

%%%%%%%%%%%%%%Table 1
\begin{table}
\tabcolsep 0.35truecm
\caption{Properties of the virialised primordial DM halos.}
\begin{tabular}{ccc} \hline
\multicolumn{1}{c}{$Mass$} &
\multicolumn{1}{c}{$T_{vir}$} &
\multicolumn{1}{c}{$R_{200}$} \\
\hline
$10^{10} M_{\odot}$ & $10^{5}~^{o}K$ & kpc \\
\hline
\hline
100 & 3.8  & 256 \\
0.1& 1.2 & 35 \\
\hline
\hline
\end{tabular}
\end{table}

The systems are then  let to  evolve until virial equilibrium is reached.
10,000  baryonic particles are then homogeneously  distributed inside the DM halo 
 with zero velocity field and low metal content ($Z~=~10^{-4}$). The initial
 gas temperature is $10^{4}$ $^{o}$K.

%%%%%%%%%%%%%%%%%%%Figure 6
\begin{figure*}
\centerline{\psfig{file=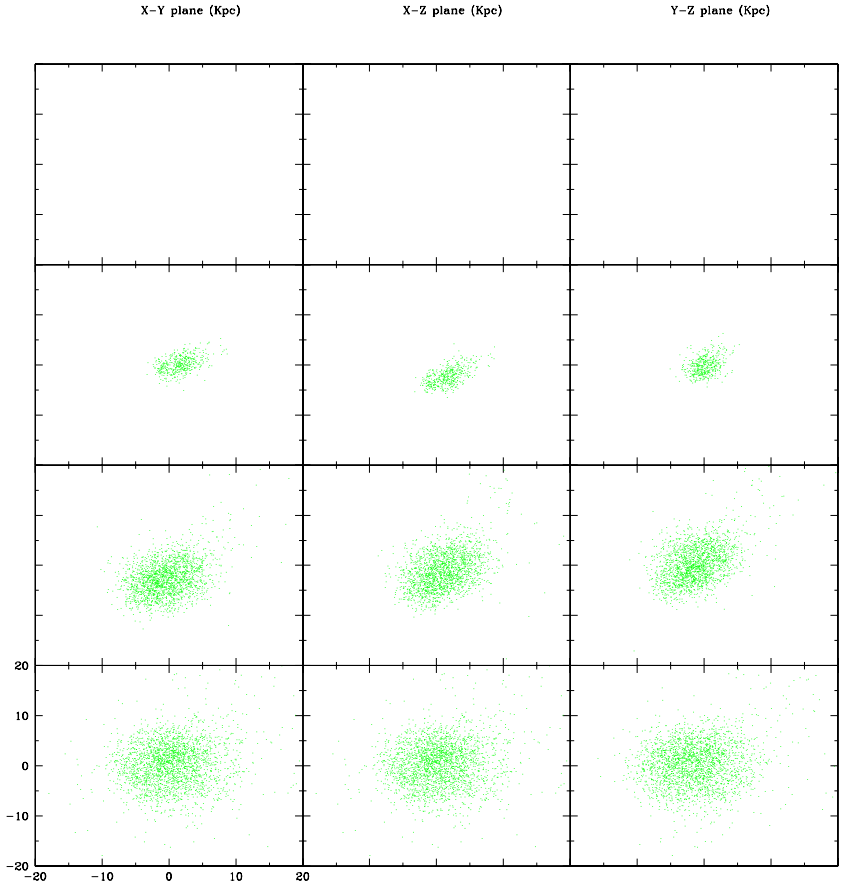,height=14cm,width=14cm}}
\caption{The formation of a dwarf elliptical. From the top to the bottom,
snapshots refer to 1, 2, 3 and 9 Gyrs.}
\end{figure*}

%%%%%%%%%%%%%%%%%%%Figure 7
\begin{figure}
\centerline{\psfig{file=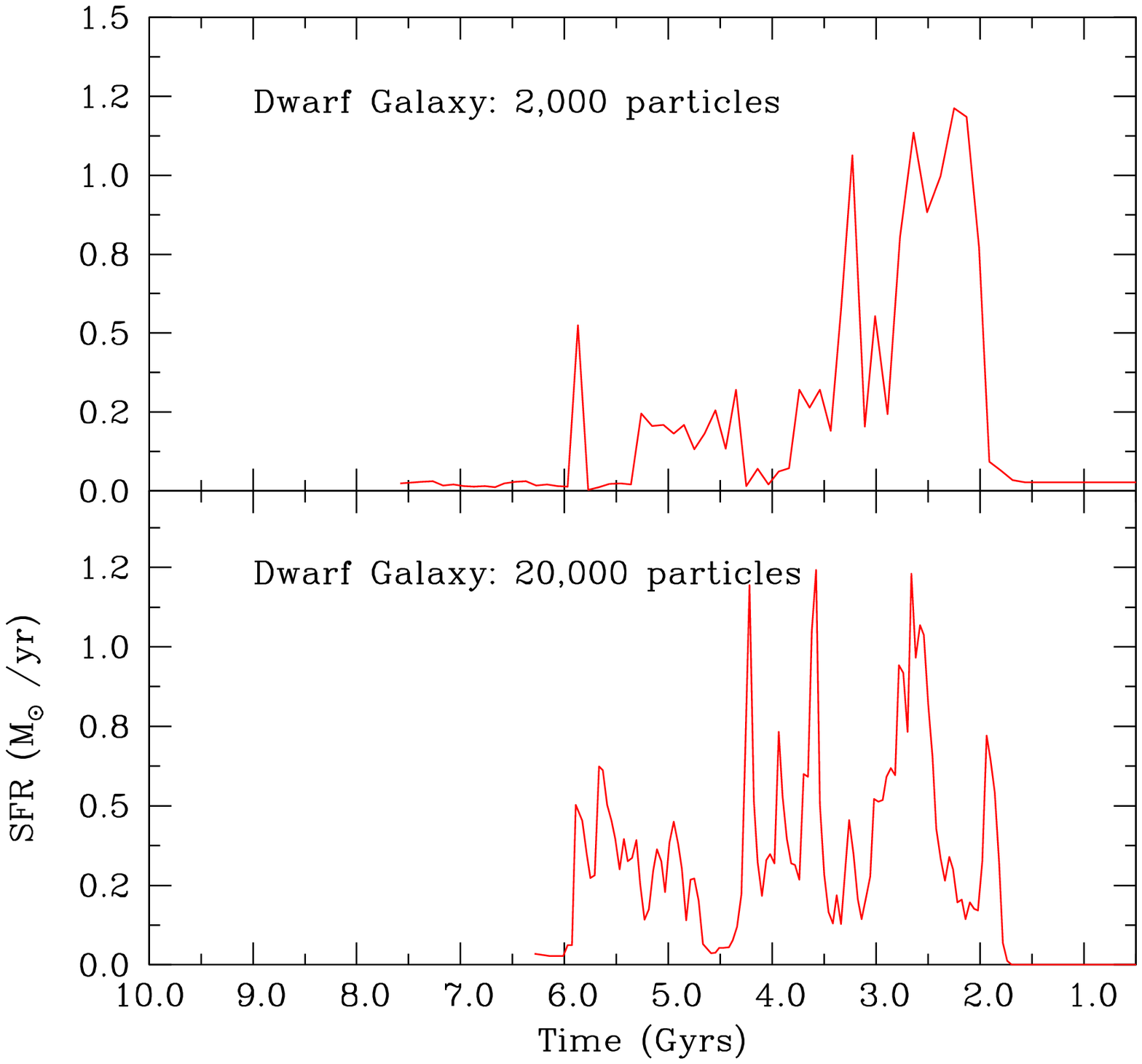,height=18cm,width=10cm}}
\caption{SF history for the dwarf elliptical as a function of time. The two
panels show the same model at increasing number of particles.}
\end{figure}

\section{A Giant Elliptical Galaxy}
In the first  simulation we set  a giant elliptical  of  gaseous mass of  $10^{11}
M_\odot$  to   form through the  collapse of a spherical 
 DM halo 
of $10^{12} M_\odot$.
Each gas particle has an initial mass of $10^{7} M_\odot$.\\
Looking at Figs.~1 and 2, baryons slowly infall towards the center of the potential well, 
gas condenses,  cools   and then finally  
stars  begin to form at  a rate of   $50 M_\odot yr^{-1}$ . The  strong episode of star
formation lasts  for about 2 Gyr and it is  marked by a large rate of production of  SN\ae~ of
type II. These, however,  are not able to expel from the galaxy a relevant fraction of 
the   gas, given its strong gravitational field.   At the end of this period, it has  been 
processed into stars  80$\%$ of the original material.   The remaining  20$\%$
is left partly in the
outermost regions of the dark halo, partly out of the virial radius.

The  collapse of  baryons in the DM potential well  can be realized by noticing that,  at 
$t=0$, in the innermost 10 kpc of the protogalaxy,  the DM halo is  10 times  denser than
the gaseous component but at 
the end of the formation of the spheroid, the stellar component  reaches a "central" density
 30 times that  of the dark component.  

Not surprisingly,  the final distribution  of  stars does not follow that of the  DM.
In fact,  as the infall proceeds,      the DM remains   $C^\infty$ scale-free, 
while the baryons develop   a    $C^1$
scale $r_e$ of  about one tenth of
the virial radius.      
More precisely, the dissipative  zero-angular momentum infall   of
the (out-to-250 kpc)  scale-free   baryonic material  produces a half-mass scale-length of
$\sim 30$ kpc.  The final stellar distribution closely
follows 
a Hernquist profile  (see Fig.~3) with effective radius $r_e$  slightly  larger   
than  that of  ellipticals of the same baryonic mass.  However,   the  simulated  value 
of $r_e$   depends on
the assumed  DM density and on the  prescriptions of star formation. Actually,
one could use the observed  $r_e$ vs  stellar mass relationship to  fine-tune   the  
semi-analytical parts of the code (Buonomo et al 2000) .\\

During the  galaxy assembly there creates  a coupling   between the baryons and the DM.
The initial halo distribution $\rho(r,0)=M_{200}/(2\pi R_{200}^2)
1/r$  gets  first  slightly contracted by the baryonic infall and then 
expanded back  by a (limited) supernovae-driven outflow. As result, the final DM distribution
is not too different from the primordial one.   
On the other side, the DM potential controls the star formation rate and efficiency, including
the fate of stellar ejecta.\\

Even in the present simplified scenario,
the  time-evolution  of the mass distribution is not that 
of an adiabatic  process (Blumenthal et al 1986). 
In fact, (see Fig.~5) the adiabatic invariant $rM(r)$ varies with time.    In
the innermost parts $r<50$ kpc,  it  increases with time while,   for larger radii,  $r<100$
kpc, it  decreases as  the collapse  proceeds. The baryonic infall develops  
shell crossings as effect of the strong radial 
dependence of the  the energy (un)balance among  cooling, heating and 
release of gravitational energy. 

%%%%%%%%%%%%%%%%%%%%Figure 8
\begin{figure}
\centerline{\psfig{file=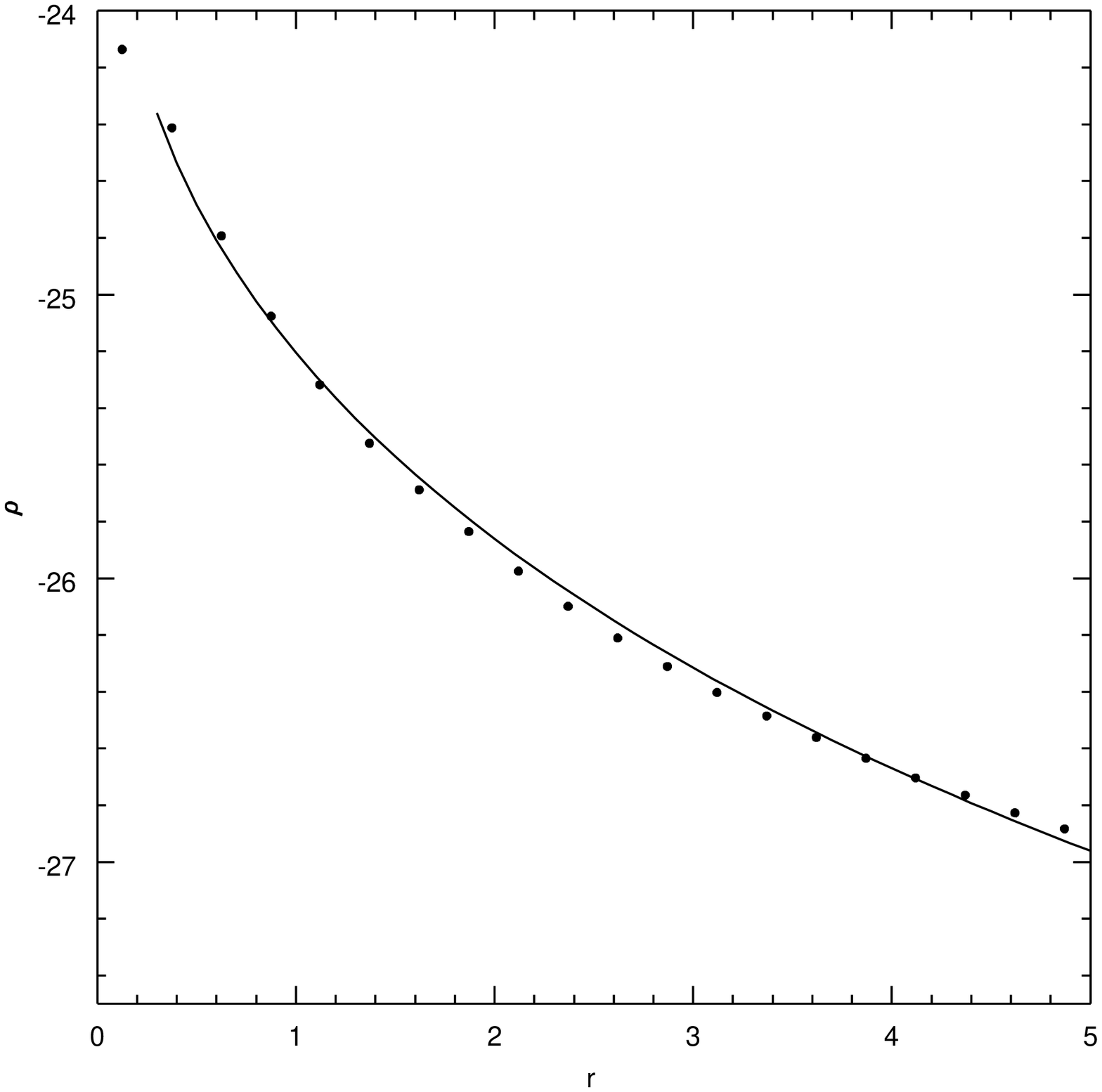,height=9cm,width=9cm}}
\caption{Baryons final density profile. Overimposed is a Hernquist profile for $r_e~=~2.8~$ kpc.}
\end{figure}

%%%%%%%%%%%%%%%%%%%%Figure 9
\begin{figure}
\centerline{\psfig{file=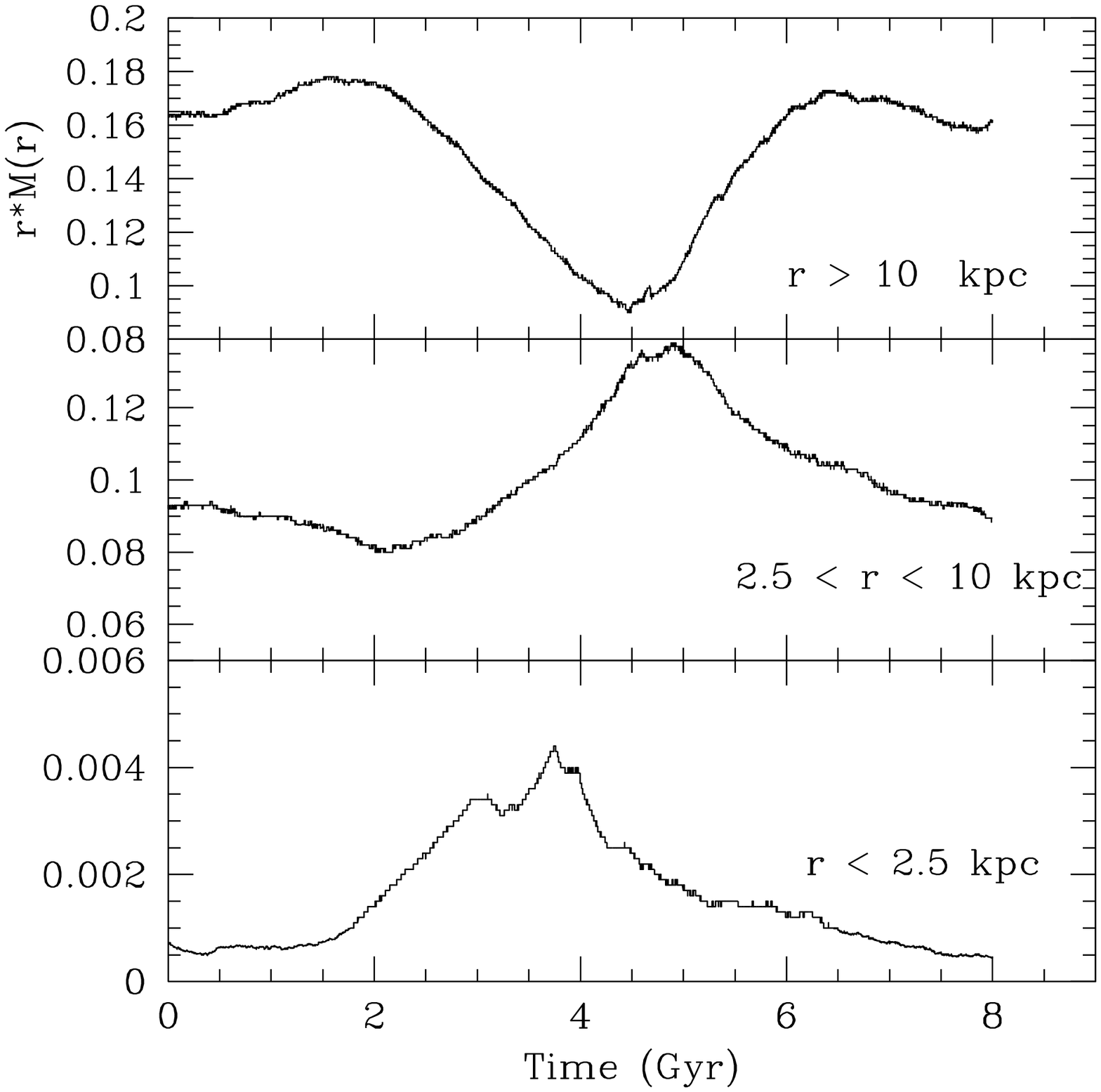,height=9cm,width=9cm}}
\caption{Adiabatic invariant  evolution for the dwarf galaxy in three different
galaxy regions. $r \times M(r)$
is in code units.}
\end{figure}

\section{A Dwarf Elliptical Galaxy}
The second simulation concerns a proto-elliptical of  gaseous mass of  $10^{8}
M_\odot$  forming  through the monolithic collapse of a spherical 
 virialized DM halo
of $M_{200} = 10^{9} M_\odot$. In this case a gas particle has an initial mass of
of $10^{4} M_\odot$.\\
Despite  the virial radius is now much smaller
($R_{200}=35$ kpc ) the gas is not more rapid in condensating,   cooling  and  forming     stars.
 Notice that  both the  free-fall time and the cooling time are  smaller.
 
The  time evolution of the SFR in
this case cannot be described as  a a single burst,  but rather as a series of several episodes,
  each one  of  about $0.5-1 M_\odot yr^{-1}$,  lasting for over  4 Gyr  (see Figs.~6 and 7),
  a result which is in close agreement with observations of Dwarf Galaxies in the Local
  Group (Mateo 1998).

 At the end of this period  it has been   processed into long-living stars only 
 $20\%$ of the
original material.   The remaining   has been  
thrown out from the galaxy or left,  unprocessed, behind,  in the outermost
regions of the DM halo.

In the central regions,
 the baryonic density increases as the collapse proceeds  to
reach "at  the  center" 30 times the 
density of the dark component.  The stellar scale-length is  $r_e \sim 2.8 $ kpc
not too different from that of  ellipticals of the same baryonic mass (see Fig.8).
Also in this case there is  a clear 
coupling between baryons and DM during the galaxy assembly.

However, in this case,  due to
the   weaker gravitational field,    most of the gas gets expelled 
from the galaxy. 
The  halo  density evolves during the  baryonic infall,  in  a non-adiabatic way, as shown 
 in Fig.~9. The final DM distribution is not too different from the primordial one
 (see Fig.~10).\\
  
%%%%%%%%%%%%%%%%%%%%Figure 10
\begin{figure*}
\centerline{\psfig{file=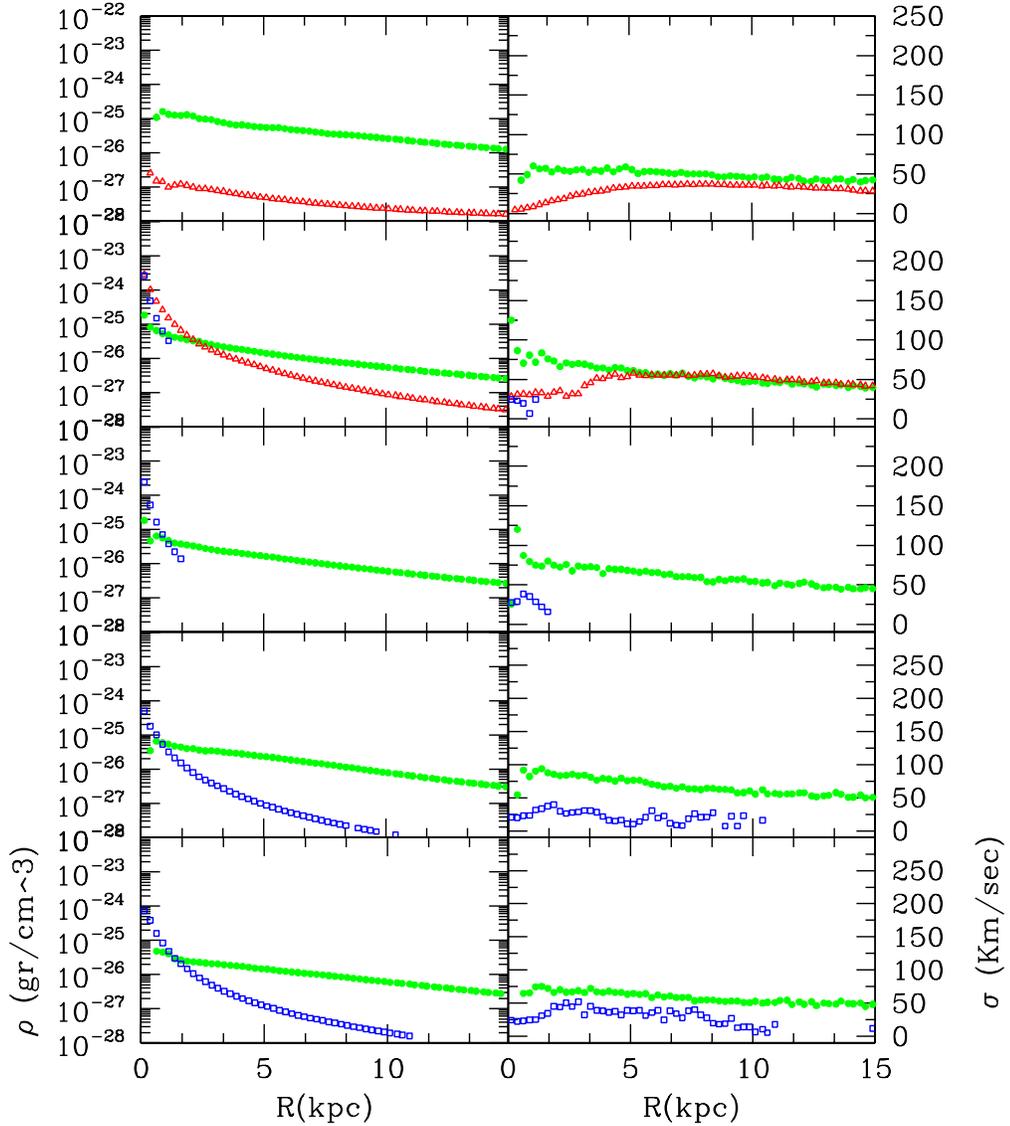,height=16cm,width=16cm}}
\caption{Evolution of density and dispersion profiles for the Dwarf
Elliptical Galaxy. Solid circles stand for DM, open triangles for
gas and open squares for stars. From the top to
the bottom profiles refer to 1.5, 2, 3.5, 5 and 9 Gyr.}
\end{figure*}

\section{Discussion and Conclusions}
In this paper we have investigated the coupling between DM and baryons during an assumed 
monolithic formation of two galaxies: a giant and a dwarf spheroidal galaxy.\\
Starting from a homogeneous distribution, baryons sink toward the center of the virialized DM halo
due to cooling instability.
The smallest object has about 10 independent episodes
of star formation $0.3 ~Gyr $ long, while the largest has basically one single burst 
10 times longer.  This difference nicely matches the available observations. 
Giant elliptical galaxies
are indeed old objects dominated by a single burst of star formation ( Bender et al 1992),
whereas dwarf ellipticals exhibit basically an irregular and intermittent SF history
(Mateo 1998). \\
A crucial and sensitive aspect of our simulations is that
both the giant and the dwarf object have been modeled with the same number of
particles, which implies an increase by a factor $10^{3}$ in the mass resolution
passing from the giant to the dwarf.\\

To check whether the results for the SF histories depend on the 
mass resolution of our simulations,
the giant galaxy has been re-simulated using 2,000, 20,000 and 200,000
particles, whilst the dwarf galaxy has been re-simulated using 2,000
and 20,000 particles.\\
This way the mass resolution goes from $10^{8}$ to $10^{6} M_{\odot}$
for the giant galaxy, and from $10^{5}$ to $10^{4} M_{\odot}$
for the dwarf galaxy.
The results are summarized in Fig.~2 and Fig.~7, where we compare the SF histories
of the two galaxy models at increasing mass resolution.\\

In both cases we show a reasonable convergence of the results.\\

As for the giant galaxy (see Fig.~2),
we show that the SF history is dominated by a single ancient burst of SF, whose
trend does not change too much passing from 2,000 to 200,000 particles.
Nevertheless a more careful analysis shows that  a difference 
emerges passing from 2,000 to 20,000 particles, whereas
passing from 20,000 to 200,000 particles does not change 
significantly the global result.\\
Although the SF peak is almost the same in all the simulations,
the low resolution run produces a burst larger than
the other two cases. Correspondingly
the amount of gas consumed (and hence the final mass in stars) passes
from $90\%$ in the low resolution run to about $80\%$ in the medium and high resolution run.
This means that at increasing number of particles the results asymptotically
converges.\\
These conclusions confirm previous analysis on the performances of the SPH method
(Steinmetz \& M\"uller 1993, Thacker et al 1999). By comparing different SPH 
implementations
they find that the minimum number of particles required to calculate
local physical variables in dynamically evolving systems is about 10,000,
and that the SPH method can give reasonable results also by using a small
number of particles.\\

On the other hand, changing the resolution by a factor 10 does not alter
the SF history of the dwarf galaxy (see Fig.~7), which exhibits several episodes of SF,
although the number and position of the peaks does not coincide exactly.\\

The basic results of this paper can be summarized as follows:

\begin{description}
\item{$\bullet$} the stars which are formed show a final distribution
much different from the DM distribution and  can be represented by
a Hernquist profile with a  length-scale $r_e$; 
\item{$\bullet$} a particular aspect of the mass distribution is that the regions inside $r_e$ , 
are  baryon dominated, while the  DM is   the main  mass component 
at  outer radii, as observed in real ellipticals;
\item {$\bullet$} finally is worth noticing that, in both objects, the final 
DM velocity dispersion is about 1-2 times
the stars velocity dispersion, in agreement with Loewenstein \& White III (1999) findings.
\end{description}

In forthcoming papers we are going to analyze simulations which adopt different initial conditions
for DM, in order to test the effect of the DM initial properties (density profile,
velocity dispersion and so forth) on the final baryon distribution.

\section{Acknowledgments*}
The authors acknowledge very useful discussion with dr. E. Pignatelli
and Prof. L. Danese. Moreover we thanks an anonymous referee for his
detailed  comments on the first version of
this paper. We acknowledge financial support from Italian Ministery
of Research, University, Science and Technology (MURST).


\begin{thebibliography}{}

\bibitem{}
Barnes J. E., Hut P., 1986, Nature 324, 446

\bibitem{}
Bender R., Burstein D., Faber S.M., 1992, ApJ 399, 462

\bibitem{}
Bertola F., Pizzella A., Persic M., Salucci P., 1993, ApJ 416, L45

\bibitem{}
Blumenthal G.R., Faber S.M., Flores R., Primack J.R. 1986, ApJ 301, 27

\bibitem{}
Buonomo F., Carraro G., Chiosi C., Lia C., 2000, MNRAS 312, 371

\bibitem{}
Carraro G., Lia C., Chiosi C., 1998, MNRAS 298, 1021

\bibitem{}
Chiosi C., Maeder A., 1986, ARA\&A, 24, 329

\bibitem{}
Cole S., Lacey C., 1997, MNRAS 281, 716

\bibitem{}
Danziger I.J., 1997,in "Dark and Visible Matter in Galaxies", 
Persic M. \& Salucci P. eds, ASP Vol. 117, p.28 

\bibitem{}
Dubinski J., Carlberg R., 1991, ApJ 378, 496

\bibitem{}
Gingold R.A., Monaghan J.J., 1977, MNRAS 181, 375

\bibitem{}
Giraud E., 2000, ApJ 531, 701

\bibitem{}
Hernquist L., 1990, ApJ 356, 359

\bibitem{}
Hernquist L., Katz N., 1999, ApJS 70, 419

\bibitem{}
Hollenback D., McKee C.F., 1979, ApJS 41, 555

\bibitem{}
Huss H., Jain B., Steinmetz M., 1999, ApJ 517, 64

\bibitem{}
Jing  Y.P., Suto Y., 2000, ApJ 529, L69

\bibitem{}
Katz N., 1992, ApJ 391, 502

\bibitem{}
Lia C., Carraro G., 2000, MNRAS 314, 145

\bibitem{}
Loewenstein M., White III R.E, 1999, ApJ 518, 50

\bibitem{}
Lucy, L., 1977, AJ 82, 1013

\bibitem{}
Mateo M., 1998, ARA\&A 36, 435

\bibitem{}
Miller G. E., Scalo J. M., ApJS 41, 513

\bibitem{}
Moore B., Governato F., Quinn T., Stadel J., Lake G., 1998, ApJ 462, 563

\bibitem{}
Navarro J.F., Frenk C.S., White S.D.M., 1996, ApJ 462, 563

\bibitem{}
Persic M., Salucci P., Stel F., 1996, MNRAS 281, 27

\bibitem{}
Press W. H., Flannery B. P., Teukolsky, S. A., Vetterling, W. T.,
{\it Numerical Recipes}, 1989, Cambridge: Cambridge University Press

\bibitem{}
Salucci P., Persic M., 1997,in "Dark and Visible Matter in Galaxies", 
Persic M. \& Salucci P. eds, ASP Vol. 117, p.1 

\bibitem{}
Steinmetz M., M\"uller E., 1993, A\&A 268, 391

\bibitem{}
Sutherland R. S., Dopita M. A., 1993, ApJS 88, 253

\bibitem{}
Syer D., White S.D.M., 1998, MNRAS 293, 337

\bibitem{}
Swaters R., 1999, PhD Thesis, Groningen University

\bibitem{}
Thacker R.J., Tittley E.R., Pearce F.R., Couchman H.M.P., Thomas P.A., 1998,
({\tt astro-ph/9809221})

\bibitem{}
Thornton K., Gaudlitz M., Janka H.-Th., Steinmetz M., 1998,
ApJ 500, 95

\bibitem{}
Van den Bosch , Robertson J.J., de Block W.J.G., 2000, ({astro-ph/9911372})

\bibitem{}
Weil  M.L., Eke V.R., Efstathiou G., 1998, MNRAS 300, 773

\end{thebibliography}
\end{document}